\documentstyle[prl,twocolumn,aps,epsf,epsfig]{revtex}
\begin{document}

\twocolumn[
\hsize\textwidth\columnwidth\hsize\csname@twocolumnfalse\endcsname
\draft

\title{The so-called two dimensional metal-insulator transition}
\author{S. Das Sarma and E. H. Hwang}
\address{Condensed Matter Theory Center, 
Department of Physics, University of Maryland, College Park,
Maryland  20742-4111 } 
\date{\today}
\maketitle

\begin{abstract}

We provide a critical perspective on the collection of low-temperature
transport phenomena in low-density two-dimensional semiconductor
systems often referred to as the 2D metal-insulator transition.
We discuss the physical mechanisms underlying the anomalous behavior
of the two-dimensional effective metallic phase and the
metal-insulator transition itself.
We argue that a key feature of the 2D MIT physics is the 
long-range bare Coulombic
disorder arising from the random distribution of charged impurities in
the low-density 2D semiconductor structures.

\noindent
PACS Number : 71.30.+h; 73.40.Qv

\end{abstract}
\vspace{0.5cm}
]

\newpage

\section{introduction}

The set of experimental observations, collectively referred to as
``the two dimensional (2D) metal-insulator transition (MIT)'', was
discovered \cite{Kravchenko94} in 1994 by Kravchenko and Pudalov in an
important series of low temperature ($T \le 1-4K$) transport
measurements in high-mobility low-density electron inversion layers in
Si metal-oxide-semiconductor-field-effect transistor (MOSFET)
structures. The original experimental observations, which have since
been qualitatively reproduced in Si-MOS systems \cite{lewalle} and in
many different 
2D semiconductor systems (e.g. p-GaAs \cite{pGaAs}, n-GaAs
\cite{lilly}, SiGe \cite{senz}, AlAs \cite{AlAs}), 
consist of a striking temperature and density dependence of the
measured 2D resistivity $\rho(T,n)$, or equivalently, the 2D
conductivity $\sigma \equiv \rho^{-1}$, at low temperatures and
densities in 2D systems of high quality or low disorder (as reflected in
high mobilities). 

One remarkable feature of the 2D MIT
transport data, which continues to attract a great deal of attention, is
the apparent existence of a ``critical carrier density'' $n_c$ which
seems to sharply distinguish the effective metallic ($n>n_c$) and the
effective insulating ($n<n_c$) phase of the 2D system at low
temperatures, with the effective `metal' or `insulator' being defined
by the temperature dependence of resistivity: $d\rho/dT >0$ (metal);
$d\rho/dT <0$ (insulator) at low temperatures ($\sim 100$ mK).
An equally remarkable  observation in the
original report \cite{Kravchenko94} is the extremely strong anomalous
``metallic'' 
(i.e., $d\rho/dT>0$) temperature dependence of the resistivity
$\rho(T)$ in the density range just above $n_c$ (approximately in the
$n_c \le n \le 3n_c$ range) where $\rho(T)$ could increase by as much
as a factor of three at a fixed density (above $n_c$) for a modest
change of temperature from $T\sim 0.1$K to $T\sim 1-4$K.
Such a huge temperature dependence of metallic resistivity at low
temperatures is completely unheard of in any nonsuperconducting
metallic systems where the resistivity in the $T\le 4$K range
essentially exhibits no temperature dependence (the so-called
Bloch-Gr\"{u}neisen regime with $\rho \sim \rho_0 + O(T^5)$) as the
acoustic phonons (the dominant source of the temperature dependence of
$\rho$ in 3D metals) become thermally frozen leading to a
low-temperature suppression of any temperature dependent scattering.
By contrast, the low-density (and high-quality, i.e., high mobility)
2D ``metallic'' systems discovered by Kravchenko and Pudalov seem to
have an {\it approximately} linear temperature dependence in the $T\sim 1$K
regime, $\rho \approx \rho_0 + O(T)$, in sharp contrast to the
``expected'' Bloch-Gr\"{u}neisen metallic behavior. (At much higher
densities, however, high-mobility and high-density 2D electron systems
manifest \cite{stormer} almost no temperature dependence similar to 3D
metallic 
behavior, showing an essentially temperature-independent saturation of
low temperature ($\le 4$K) resistivity which is consistent with the 2D
Bloch-Gr\"{u}neisen behavior.) It should, however, be mentioned that
at low temperatures ($\le 50$mK), $\rho(T)$ saturates in 2D MIT systems
also, but it is unclear whether this is intrinsic or an electron
heating effect. An early review of the basic 2D MIT phenomena can be
found in ref. \onlinecite{d2}, but much of our current theoretical
understanding of the subject has happened more recently.

The observation of a sharp density dependent 2D MIT (with the system
being an effective metal for $n >n_c$ and an effective insulator for
$n<n_c$) and the associated unusually strong temperature dependence of
the metallic resistivity for $n \ge n_c$ (but {\it not}
for $n\gg n_c$) was immediately greeted by a substantial fraction
of the community as the discovery of a new (and perhaps quite exotic)
`metallic' phase stabilized by the strong electron-electron interactions
in the low-density 2D system. 
The case for the 2D ``metallic'' phase being exotic was further
reinforced by the anomalously strong temperature dependence of the
effectively metallic 2D phase.
The reason for this enthusiasm about 2D
MIT phenomena is the theoretical understanding developed in the late
70s and the early 80s, going by the topical names of ``weak
localization'' and/or ``scaling theory of localization''
\cite{d2,weak_local,rmp}, which 
asserted on firm grounds that a disordered noninteracting 2D
electron system at $T=0$ is strictly a localized Anderson insulator
(no matter how weak the disorder may be -- only the localization
length depends exponentially
on the strength of disorder), and as such there is no
$T=0$ 2D metal in the thermodynamic limit \cite{rmp}. If the 2D MIT
phenomenon indicates the existence of a true (rather than an
effective) 2D metallic phase with a density-driven ($T=0$) quantum
phase transition between the $n>n_c$ 2D metal and the $n<n_c$ 2D
insulator, then it is a direct violation of the scaling theory of
localization which states that there can be no true 2D metallic phase,
at least for noninteracting electron systems.

It was soon realized \cite{d2}
that if the 2D MIT phenomenon is a true quantum
phase transition with the high-density ($n>n_c$) phase being a real
metal, then much of our understanding of two dimensional electron
systems will have to be revised since the weak localization arguments
predict the noninteracting 2D system to be an insulator, and
therefore, if the interacting 2D system is indeed a true 2D metal,
then the two systems (the noninteracting insulator and the interacting
metal) cannot be adiabatically connected and the one-to-one
correspondence between the noninteracting and the interacting system,
which is at the heart of the Landau Fermi liquid theory, would fail
for a disordered interacting 2D system. This would be extremely
dramatic and important since an ideal (i.e., no disorder) interacting
2D system is known on rather firm grounds to be a Fermi liquid. There
has therefore been a great deal of activity focusing on the important
question of whether the 2D MIT is a true $T=0$ quantum phase
transition (between an insulator and a metal) or a sharp crossover
between a weak and strong localized phase (i.e. from an effective
finite-temperature
metal at $n\gg n_c$ to an insulator at $n <n_c$).

Another physical effect attracting considerable attention
\cite{si1,si2,pga1,pga2,nga}  is
the application of an in-plane parallel magnetic field on the 2D
metallic phase and on the 2D MIT itself. A fairly modest ($1-10T$
range) in-plane magnetic field parallel to the 2D electron layer gives
rise to a rather large positive magneto-resistance at low temperatures
and carrier densities. Although the quantitative details of this
magneto-resistance are somewhat system specific, the effect could be
quite spectacular for densities just above the zero-field critical
density for the 2D MIT where the parallel field, at least in Si
MOSFETs, could drive the nominal zero-field effective metallic phase
into a finite field insulating phase, consequently producing an
enormous magnetoresistance. This means that the critical density has
an apparent field dependence with $n_c(B) >n_c(B=0)$. This has been
referred to as the field-induced destruction of the 2D metallic
phase. There are also claims of the magnetic field induced (as
distinct from density induced) quantum phase transition in 2D MIT. The
generic effect of the parallel field in the low-disorder, low-density
2D system seems to be the large magnetoresistance effect, which has
been seen in several different low-density 2D systems of interest.

Following this brief and necessarily sketchy introduction to the 2D
MIT phenomena and the properties of the anomalous 2D metallic phase,
we provide, in the rest of this review article, a theoretical
phenomenological perspective on the 2D MIT phenomena concentrating on
the broad and fundamental qualitative issues of the greatest
importance. We do not make any attempts at reviewing the whole 2D MIT
literature, which is now pretty vast exceeding 500 publications over
the last 10 years. Our goal is to bring the reader of this review up
to date on our qualitative understanding of the 2D MIT phenomena
without getting bogged down in the specific quantitative details of
different 2D systems. Therefore, only selected 
publications of direct relevance to this review article are cited.
An early comprehensive review of 2D MIT exists in the literature
\cite{d2}.

\section{Background}

The hint for an anomalously strong metallic temperature dependence in
2D carrier systems was already present in the early 1980s \cite{cham}.
Even a cursory look at the $\rho(T)$ data in the 1994
Kravchenko {\it et al.} paper at various carrier densities explicitly
shows that the strong temperature dependence of $\rho(T)$ starts deep
in the metallic phase at densities $n \gg n_c$ with the temperature
dependence of $\rho(T)$ becoming stronger with decreasing
density as the density approaches $n_c$ from above. Thus the signature
of the anomalous metallic conductivity in the sense of a
strong variation in $\rho(T)$ with a modest variation in temperature
already exists at high densities, evolving continuously and
monotonically to the anomalously strong metallic behavior observed by
Kravchenko {\it et al}. in 1994. Indeed the first observation of such
an anomalously strong temperature dependence of 2D resistivity was
made by Cham and Wheeler \cite{cham} in 1980 who found that a
high-mobility Si MOSFET manifests a strong and approximately linear
temperature dependence at low temperatures ($1-4$K). This observation
of `strong metallicity' in Si MOSFETs was followed up \cite{smith} by
a number 
of other experimental groups throughout the 1980s eventually
culminating in the Kravchenko {\it et al}. observations in 1994 which,
by virtue of using extremely high-mobility Si MOSFETs and lower
measurement temperatures ($0.3 - 4$K), found a much more dramatic
temperature dependence in the resistivity, namely roughly a factor of
three change in $\rho(T)$ in the temperature range of $0.3-4$K whereas
the earlier measurements, which could only investigate higher carrier
densities since the sample quality was not that good (and consequently
$n_c$ was rather high), found more like a 20\% (or less) 
temperature induced
change in $\rho(T)$. But at least some of 
the salient features, albeit on a less
dramatic quantitative scale, were already present in the experiments
\cite{cham,smith} dating back to 1980s. Indeed a very recent
publication \cite{pudalov} in 
the 2D MIT literature presents data rather similar to the original
Cham and Wheeler data, claiming new
understanding of the 2D metallic phase,
ignoring the early literature.

Just as the experimental observation of an effective 2D metallic phase
with a strongly temperature dependent $\rho(T)$ dates back to 1980, so
does its qualitative theoretical understanding. Stern pointed out,
already in 1980 in a companion paper \cite{stern} to the Cham and Wheeler
paper, that an ``unexpectedly'' strong almost-linear temperature
dependence of $\rho(T)$ would arise in 2D systems (even for $T/T_F \ll
1$) due to the peculiar nature of the 2D screening function which, at
$T=0$, has a cusp at wave vector $q=2k_F$. Since $2k_F$ scattering is
the most dominant resistive scattering at low temperatures, thermal
smearing of this $2k_F$ cusp (i.e., the Kohn anomaly) would lead to
rather strong temperature dependence of resistivity. Later on, it was
explicitly shown \cite{gold} within this screening model that the
leading-order temperature correction to $\rho(T)$ is indeed linear in
$T/T_F$ as Stern found in his numerical calculations. (We note that
the temperature correction in these early experiments was only
$10-20\%$ whereas the current 2D metallic phase often 
manifest large temperature correction exceeding 100\%.)

Thus both the experimental discovery and the theoretical understanding
of the anomalous 2D metallic phases, with $\rho(T)$ showing strong
(and approximately linear) temperature dependence dates back to 1980
although much of the current 2D community seems to be completely
ignorant of this pre-history.

Of course the post-1994 2D MIT experiments, starting with the
pioneering papers of Kravchenko {\it et al.}, are spectacular in the
strongly anomalous metallic temperature dependence (often by a factor
$2-4$) of $\rho(T,n)$ and in the sharp metal-insulator ``transition''
at $n=n_c$ with 
$d\rho/dT >0$ for $n<n_c$ and $d\rho/dT <0$ for $n<n_c$. Also the
parallel field effect results are all new (in fact, there has even
been at least one tantalizing early observation of an anomalous
parallel magnetic field suppression of 2D conductivity reported in
ref. \onlinecite{dolgopolov}), just as the vast literature on 2D MIT
findings in many different semiconductor materials and systems also
are. But the basic observation of a reasonably strong low-temperature
``metallic'' behavior, with $\rho(T)$ increasing strongly with temperature,
certainly dates back to the work of Cham and Wheeler in 1980.
Also, the observation of a metal-insulator transition in 2D systems
was fairly routine \cite{Ando_rmp} in the 1970s with decreasing
carrier density --- the new feature in the current 2D MIT phenomena
being the strong temperature dependence of $\rho(T)$ in the putative
metallic phase for $n>n_c$. Thus the really key new feature of the
current 2D MIT literature is the strong metallic temperature
dependence for $n\ge n_c$, which we believe to be the fundamental key
ingredient of the 2D MIT phenomena. This anomalous 2D {\it
  metallicity} (i.e. the strong metallic temperature dependence of
$\rho(T)$ for $n>n_c$) is the primary focus of our perspective.

\begin{figure}
\epsfysize=5.2in
\centerline{\epsffile{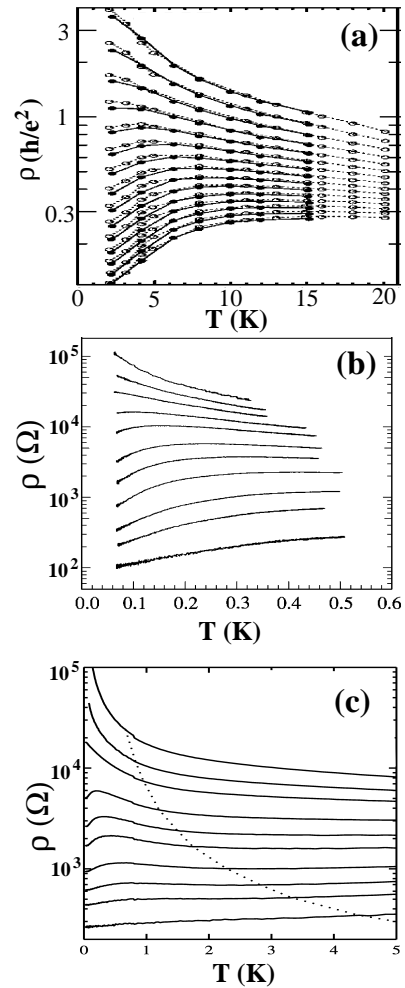}}
\vspace{0.5cm}
\caption{
Experimental $\rho(T)$ over a range of density for three different
2D systems: (a) 
Si-MOSFET (where $n_c =1.0\times 10^{11} cm^{-2}$) [2]; (b) p-GaAs
(where $n_c = 4\times 10^{9} cm^{-2}$) [41]; and (c) n-GaAs (where $n_c
= 2.3 \times 10^9 cm^{-2}$) [4]. In Fig. 1(a) the
density ranges 
from 8.9 to 22.4$\times 10^{11} cm^{-2}$ (top to bottom); in (b) from
0.15 to 3.2$\times 10^{10}cm^{-2}$; and in (c) from 0.16 to
1.06$\times 10^{10}cm^{-2}$. The metal-insulator transition happens in
each case at the density value where $d\rho/dT$ changes its {\it sign}
at low temperatures. The high temperature change of sign in $d\rho/dT$
(i.e., the non-monotonicity in $\rho(T)$), particularly at lower
densities in the metallic phase, arises from the
``quantum-classical'' crossover mechanism [22,23].
}
\label{fig1}
\end{figure}

\section{A qualitative explanation of 2D metallicity}

The anomalously strong metallic ($d\rho/dT>0$) temperature dependence is the
hallmark of the 2D effective metallic phase. The temperature
dependence (typically in the $T \approx 50mK - 4K$ range depending on
the material and the system) could be as large as a maximum
temperature induced relative change in the resistivity by $\sim 300
\%$ (Si MOSFETs) \cite{Kravchenko94,lewalle} to $\sim 25 \%$ (n-GaAS)
\cite{lilly}, but the  carrier density and
the temperature range over which $\rho(T)$ shows strong metallicity
depend strongly on the 2D material involved (see Fig. 1). 
The temperature dependence of
$\rho(T,n)$ at the lowest temperatures is also continuous in the sense
that it increases monotonically at the lowest temperature with
decreasing carrier density until the system goes over to the
insulating phase. The magnitude of the anomalous temperature
dependence certainly depends strongly on the material involved and is
not just a function of the carrier density or the dimensionless density
parameter (the Wigner-Seits radius)
$r_s = (\pi n)^{-1/2}/a_B$ with $a_B=\kappa \hbar^2/me^2$, where $r_s$,
the average inter-electron separation measured in the effective
Bohr radius $a_B$, is the interaction parameter giving the ratio of
the average Coulomb electron-electron potential energy to the
noninteracting kinetic energy.

It was suggested by us \cite{DH1} in 1999 that the anomalously strong
metallic temperature dependence discovered by Kravchenko {\it et al.}
arises from the physical mechanism of temperature, density, and wave
vector dependent screening of charged impurity scattering in 2D
semiconductor structures, leading to a strongly temperature dependent
effective quenched disorder controlling $\rho(T,n)$ at low
temperatures and densities. In Fig. 2 we show our calculated
resistivity for different parameters and systems within the screening
theory \cite{DH1,DH2,DH3,DH4,DH5}.
This is the same screening mechanism invoked originally by Stern in
1980 in the context of the $\rho(T,n)$ measurements by Cham and
Wheeler except now the effect is strongly enhanced at very low carrier
densities and temperatures achievable in the high-quality 2D samples.

We have explicitly shown \cite{DH1,DH2,DH3,DH4,DH5} that the
requirements for the  observation of a large
temperature-induced change in resistivity are the following: (1) A
comparatively large change in the value of the dimensionless
temperature $t=T/T_F$ which can be accomplished by having a $T_F$
which is at most a few degrees Kelvin so that the temperature regime
$0 \le T \le T_F$ can be explored in transport studies without phonon
scattering complications becoming important; (2) the low value of
$T_F$ must be lower than the effective phonon scattering temperature
$T_{ph}$ above which phonons start contributing to the temperature
dependent resistivity -- $T_{ph}$ turns out to be a strong function of
the semiconductor material and carrier density, and could be as low as
$0.5-1K$ in GaAs 
systems and as high as 10K or above in Si MOSFETs;
(3) the strong screening condition, $q_{TF} \gg 2k_F$, must be
satisfied in order to see a large temperature induced change in
$\rho(T)$; (4) sufficiently high quality or low disorder in the system
must be achieved (i.e., low $n_c$) so that the low $T_F$ condition can
be attained.
The conditions (1) and (3) above explicitly necessitate a 2D system
with effectively low carrier density for the observation of 2D
metallicity whereas the condition (4) explicitly requires high quality
or low disorder.


\begin{figure}
\epsfysize=5.5in
\centerline{\epsffile{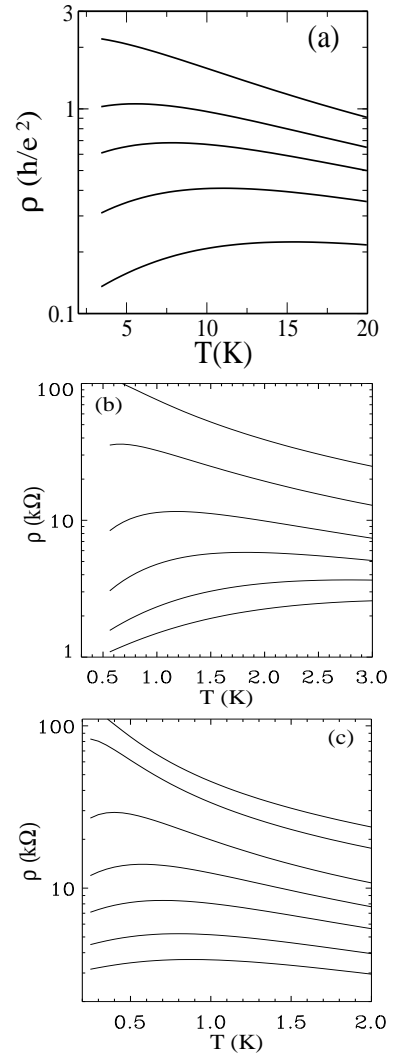}}
\vspace{0.5cm}
\caption{
Calculated $\rho(T)$ over a range of density for three different
2D systems: (a) 
Si-MOSFET for $n =$8, 10, 12, 15, 20$\times
10^{10}cm^{-2}$; 
(b) p-GaAs $n =$0.1, 0.2, 0.5, 1.0, 2.0, 3.0$\times
10^{10}cm^{-2}$; 
and (c) n-GaAs for $n =$0.4, 0.5, 1.0, 2.0, 3.0, 4.0, 5.0$\times
10^{10}cm^{-2}$. Both the anomalously strong temperature dependence in
the metallic phase and the non-monotonicity of $\rho(T)$ are evident
in these theoretical results.
}
\end{figure}


The new development in 2D physics has been the availability of
high-quality or equivalently low-disorder samples, starting with the
pioneering work of Kravchenko {\it et al.}, where all these conditions
can simultaneously be satisfied. The old Si MOSFETs, where Cham and
Wheeler made the first experimental observation of the reasonably
strong temperature dependence in $\rho(T)$, had maximum mobilities of
the order of $10^3 cm^2/Vs$ so that the low density (and low $T_F$)
regime could not really be explored without strong localization
setting in. Therefore, Cham and Wheeler were restricted to studying
metallic 
temperature dependence of $\rho(T;n)$ at carrier densities
$10^{12} cm^{-2}$ or above, where $T_F \ge 75$K in Si MOSFETs, so that
the condition of $t\equiv T/T_F$ being large without significant
phonon scattering effects simply could not be satisfied. The new Si
MOSFET samples in the post Kravchenko (i.e., after 1994) era have
maximum mobilities which are $2-5\times 10^4 cm^2/Vs$ with
consequently rather low critical densities for the strong localization
``transition'' (actually, a crossover) allowing $T_F$ to be down to a
few K so that $T/T_F$ can be large without phonon scattering becoming
significant.
This is the basic underlying reason for the observed 2D ``metallicity''.

In n-GaAs 2D structures, of course, extremely high mobilities
surpassing $10^7cm^2/Vs$ have been achieved. But, until very recently
\cite{lilly}, no strong metallic temperature dependence can be seen in
high-mobility 2D n-GaAs systems at low temperatures (i.e., in the
Bloch-Gr\"{u}neisen range where phonons are not operational) because
of two reasons: (1) The strong screening condition $q_{TF}/2k_F \gg 1$
cannot be satisfied in 2D GaAs electron systems except at rather low
carrier densities ($<10^{10}cm^{-2}$); and (2) the Fermi temperature
is relatively large in n-GaAs systems making it difficult to satisfy
the $t=T/T_F \sim 1$ criterion without having considerable phonon
scattering. This is the reason why the strong anomalous metallicity
(i.e., a large temperature induced change in resistivity) is difficult
to see in GaAs-based 2D electron systems. Recently, however, extremely
high-quality gated 2D n-GaAs systems have been fabricated \cite{lilly}
where 
the carrier density can be lowered to a remarkably low value of $10^9
cm^{-2}$. In these very special 2D n-GaAs systems, the strong metallic
temperature dependence of $\rho(T)$ shows up \cite{lilly} rather
strikingly in 
the 30mK---1K temperature range although there are complications
arising from phonon scattering that need to be taken into account.

To get a quantitative feeling we write down expressions for 2D $T_F$
and $q_{TF}/2k_F$ below:
\begin{equation}
T_F = 2\pi \hbar^2 n/(g_s g_{\nu}m k_B),
\end{equation}
where $g_s$ ($g_{\nu}$) is the spin (valley) degeneracy factor ($g_s =
2$ (2), $g_{\nu} = 1$ (2) usually for spin unpolarized GaAs (Si) systems);
\begin{equation}
\frac{q_{TF}}{2k_F} = \frac{e^2}{\hbar^2}
\frac{(g_sg_{\nu})^{3/2}m}{\sqrt{4\pi n} \kappa} \sim
\frac{(g_sg_{\nu})^{3/2}m}{\kappa n^{1/2}},
\end{equation}
where $\kappa$ is the background lattice dielectric constant. For the
purpose of comparison we also write down the corresponding
$q_{TF}/2k_F$ in 3D systems:
\begin{eqnarray}
\left ( \frac{q_{TF}}{2k_F} \right )_{3D}& =& \frac{(g_s
  g_{\nu})^{2/3}}{n^{1/6}} \left (
  \frac{me^2}{\kappa \hbar^2} \right )^{1/2} \frac{1}{\pi^{5/6}6^{1/6}2^{1/2}}
  \nonumber \\
  & \sim & \frac{(g_sg_{\nu})^{2/3}m^{1/2}}{\kappa^{1/2}n^{1/6}}.
\end{eqnarray}
For the sake of convenience we also write down $T_F$ and $q_{TF}/2k_F$
specifically in 2D Si MOS and n-GaAs systems: $T_F = 7.3 \tilde{n}_{Si}$ K
($42 \tilde{n}_{GaAs}$ K), where $\tilde{n}_{Si} = n/10^{11}$ and
$\tilde{n}_{GaAs} 
= n/10^{11}$ (i.e., $\tilde{n} \equiv n/10^{11}$). Thus, $T_F = 7.3K$ in
Si for $n=10^{11}cm^{-2}$ whereas $T_F=42 K$ in n-GaAs for
$n=10^{11}cm^{-2}$. Similarly, we can write $(q_{TF}/2k_F)_{Si}
\approx 11/\sqrt{\tilde{n}}$ and  $(q_{TF}/2k_F)_{GaAs} \approx
1.3/\sqrt{\tilde{n}}$. Finally, we note that $(q_{TF}/2k_F)_{2D} \propto
g_{\nu}^{3/2}r_s$ and $(T/T_F)_{2D} \propto g_{\nu}r_s^2$, whereas
$(q_{TF}/2k_F)_{3D} \propto g_{\nu}^{2/3} \sqrt{r_s}$ and
$(T/T_F)_{3D} \propto g_{\nu}r_s^2$ in terms of the valley degeneracy
factor ($g_{\nu} =2$ for Si MOSFETs and 1 for GaAs systems) and the
dimensionless interaction parameter $r_s$.

The examination of the expressions for $q_{TF}/2k_F$ and $T/T_F$
immediately reveals that the Si MOS 2D electron system is
substantially more ``metallic'' than the GaAs 2D electron system, as
is experimentally observed, at the same carrier density since
$(q_{TF}/2k_F)_{Si} \approx 10 (q_{TF}/2k_F)_{GaAs}$ at similar
density. In fact, a 2D GaAs electron system would require a factor of
100 lower carrier density than the 2D Si electron system for the two
to have ``similar'' metallicity (for example, equivalently strong
temperature dependent resistivity when expressed in terms of the
dimensionless temperature dependence $t \equiv T/T_F$) -- the
``metallicity'' is even more lopsided in favor of Si MOSFETs when one
takes into account the fact that $T_F^{GaAs} \approx 6 T_F^{Si}$ at
the same carrier density! Thus both the density and the materials
dependence of ``metallicity'' (i.e., why the strength of the metallic
temperature dependence of resistivity is enhanced with decreasing
carrier density in a given sample and why there is substantial
variation in the metallicity strength among different materials and
systems) can be qualitatively understood as arising from the
temperature dependence of effective (i.e., screened) disorder in the
system as controlled by the dimensionless parameters $q_{TF}/2k_F$ and
$T/T_F$. We note that a simple comparison of these parameters between
2D and 3D systems also immediately provides an explanation for why
such a strong metallicity is highly unlikely (but not theoretically
impossible) to occur in 3D. In particular, $(q_{TF}/2k_F)_{3D} \sim
n^{-1/6}$ in contrast to $(q_{TF}/2k_F)_{2D} \sim n^{-1/2}$, and
therefore the effective disorder is much more weakly density dependent
in 3D than in 2D, making it much more difficult to control
``metallicity'' by changing carrier density in 3D systems (e.g., a
doped semiconductor system). In addition, in 3D metals $q_{TF}/2k_F
\sim 1$ and $T/T_F \sim O(10^{-4})$ at low temperatures, so that
screening-induced temperature dependence of transport properties
(through the effective temperature dependent disorder) is not
observable in 3D metallic systems.
Also, the $2k_F$ Kohn anomaly is much sharper in 2D (a cusp, see
Fig. 3) than in 
3D, contributing to an effectively stronger 2D temperature dependence.
Thus, anomalous metallicity is highly unlikely, but not theoretically
impossible, to occur in low density 3D metallic systems.

The typical $q_{TF}/2k_F$ values in the 1994 Kravchenko experiment
were in the 8--12 range whereas it was in the 3--4 range for the Cham
and Wheeler samples explaining

\begin{figure}
\epsfysize=2.2in
\centerline{\epsffile{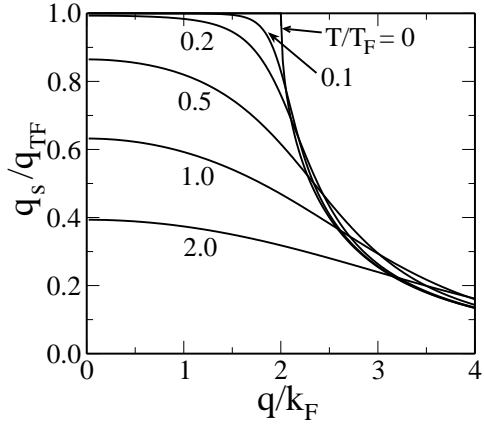}}
\vspace{0.5cm}
\caption{
The 2D screening function $q_s(q,T)$ (in units of the long-wavelength
Thomas-Fermi screening constant $q_{TF}$) as a function of dimensionless
wave vector $q/k_F$, where $k_F$ is the Fermi wave vector, for several
different 
temperatures $T/T_F=$0, 0.1, 0.2, 0.5, 1.0, 2.0 (top to bottom).  The strong
temperature-induced suppression of the $2k_F$ Kohn anomaly in screening is
evident in the figure even for very low $T/T_F$.  In the strong screening
limit the low-temperature resistivity is approximately inversely
proportional to the square of [$q_s(2k_F)/2k_F$] if charged impurity
scattering is dominant in the system, which, as is obvious from this figure,
will manifest very strong temperature dependence leading to the anomalous
metallicity at low carrier densities.
}
\end{figure}

\noindent
the much stronger temperature
dependence in the Kravchenko experiment. In addition, the $T/T_F$
values ranged in 0.01--0.05 range for the Cham and Wheeler
measurements whereas it is 0.1--1 range in the Kravchenko experiments,
again emphasizing the occurrence of
much stronger temperature dependence in the
high-mobility low-density systems. In fact, the reason for requiring
high-quality and low-disorder 2D systems for the observation of 2D
metallicity is obvious in the 
screening theory -- one must be able to
achieve fairly low carrier densities without going into a strongly
localized phases (i.e. $n_c$ should be ``low'') so that high
$q_{TF}/2k_F$ values can be attained in order to observe the strong
temperature-induced variation in the effective disorder as reflected
in the strong temperature dependence of $\rho(T)$. In systems with high
disorder, $q_{TF}/2k_F$ values cannot be large 
since low carrier densities satisfying $n >n_c$ cannot be achieved,
and therefore the
strong screening condition needed for metallicity cannot be satisfied.
The low disorder pushes the strong localization crossover
density $n_c$ to lower values enabling the $q_{TF}/2k_F$ (and $T/T_F$)
values of the 2D system to be much larger allowing the striking
observation of the strong temperature (and magnetic field) dependence
of the 2D resistivity in the effective metallic phase.
In Fig. 4 we show the $q_{TF}/2k_F$ and $T/T_F$ 
dependence of $\rho(T)$ in the screening theory.

The screening theory also provides a natural explanation for the
observed nonmonotonicity in $\rho(T)$ at larger


\begin{figure}
\epsfysize=2.in
\centerline{\epsffile{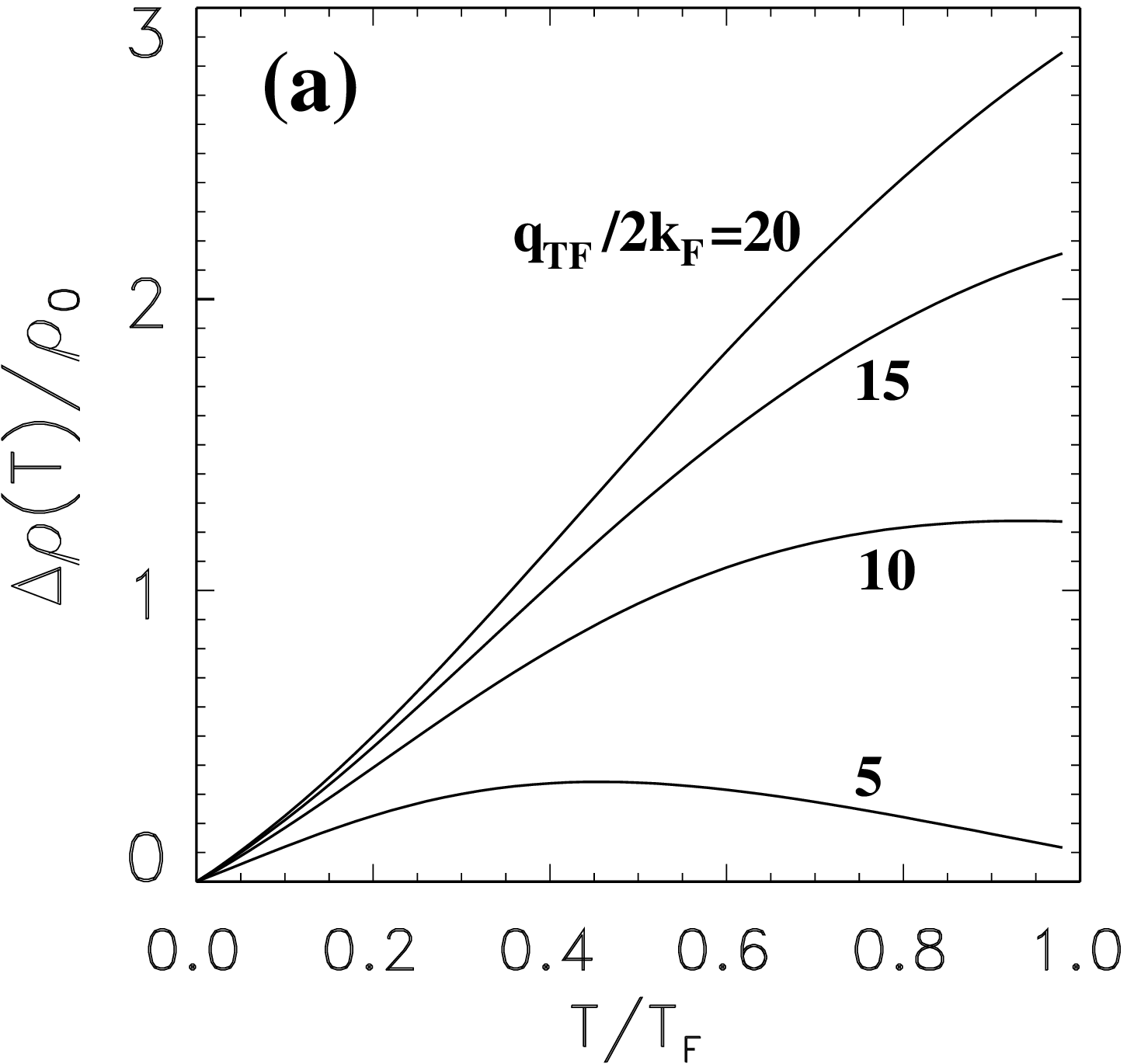}}
\epsfysize=2.1in
\centerline{\epsffile{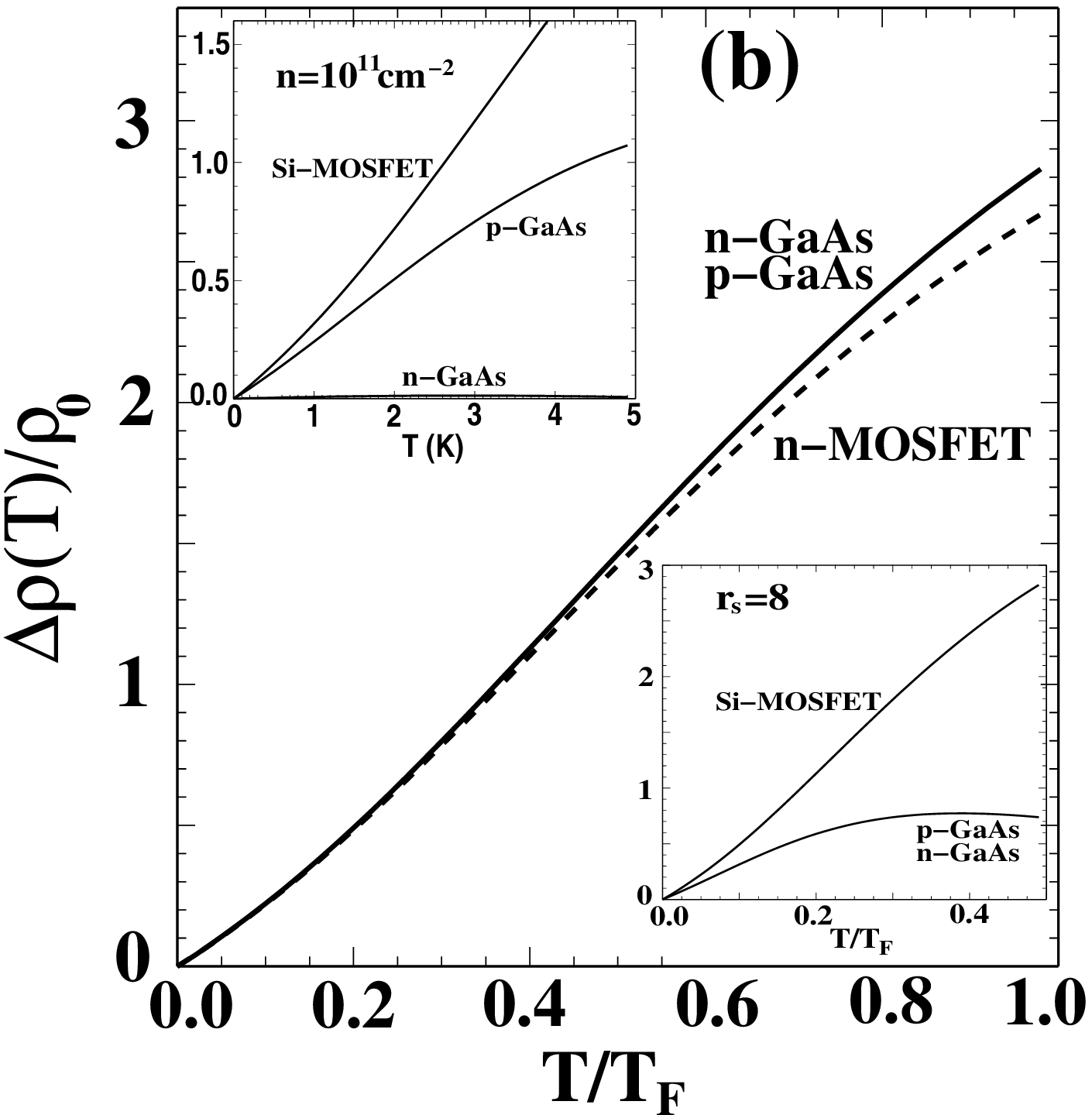}}
\vspace{0.5cm}
\caption{
(a) Calculated $\Delta \rho = \rho(T,n)-\rho_0$,
  where $\rho_0=\rho(T=0)$, for various
$q_{\rm TF}/2k_F$ as a function of $t=T/T_F$ for Si MOSFET.
(b) Calculated $\Delta \rho = \rho(T,n)-\rho_0$ as a function of 
$T/T_F$ for Si-MOSFET, p-GaAs, and n-GaAs with a fixed $q_{TF}/2k_F=15$.
In upper inset  the comparison of the
metallicity for fixed density $n=10^{11}cm^{-2}$ is given as a
function of a temperature. 
In lower inset we show the change of resistivity for fixed value of
$r_s=8$ as a function of $t=T/T_F$. 
}
\end{figure}

\noindent
temperatures ($T\ge
0.1T_F$), where $\rho(T)$ seems to slowly decrease with increasing
temperature. This arises from the ``quantum-classical'' crossover
mechanism \cite{DH1,DH2,DH5} where, for $T/T_F\sim 1$, the
resistivity goes 
down with increasing temperature due to the ``higher carrier velocity''
at higher temperatures which becomes increasingly important at higher
temperatures. The characteristic temperature scale where this
nonmonotonic crossover from 
an increasing $\rho(T)$ to a decreasing
one occurs is a strong function of the system and the carrier density
involved, and could be very low ($T/T_F \sim 0.1$) for the 2D n-GaAs
system particularly 
at higher densities.

The strong parallel magnetic field dependence of resistivity also
arises (at least partially) from the screening effect \cite{gold_b}. In
particular, the screened effective disorder would have a strong
magnetic field dependence through the spin polarization effect since a
fully spin-polarized system has weaker screening than the paramagnetic
unpolarized system as the density of states is a factor of 2 lower in
the fully spin-polarized system, i.e., $g_s =1$(2) in the spin
(un)polarized system. This means that the effective screening
parameter $q_{TF}/2k_F$ changes by a factor of 2 as the parallel
magnetic field increases from $B=0$ to $B=B_s$ where $B_s$, given by
$g\mu_BB_s =E_F$ with $g$ as the material-dependent Land\'{e}
$g$-factor and $\mu_B$ the Bohr magneton, is the saturation field
needed to fully spin-polarize the system. At the low carrier densities of
interest in the 2D MIT phenomena $B_s \sim 1-10T$, and therefore the
applied parallel field should strongly influence 2D carrier transport
through the same screening mechanism that gives rise to the
temperature dependent 2D resistivity. The strong qualitative
similarity between the temperature and the magnetic field dependence of
resistivity in Si MOSFETs has been phenomenologically discussed in the
literature \cite{pudalov1}. In the screening model, both dependences
arise from the 
weakening of screening (and consequently an enhancement of the effective
disorder) -- in one case by increasing temperature and in the other
case by increasing spin-polarization (due to the
applied parallel field). One immediate consequence of the
parallel field induced screening effect is that it predicts that
$\rho(B)$ would increase {\it only} in the $0<B<B_s$ range, and will
saturate for $B\ge B_s$ since the spin polarization effect saturates at
$B_s$. This is indeed approximately the case in Si MOSFETs where the
screening effect is the dominant mechanism in determining the
parallel magnetic field dependence.

In n-GaAs (as well as p-GaAs) 2D systems spin-polarization induced
screening modification is only a partial transport contribution of the
parallel magnetic field. Another important effect of the applied
parallel field in 2D systems is the direct orbital coupling \cite{DH6}
of the 
parallel field to the 2D carriers by virtue of the finite layer width
of the {\it quasi}-2D electron systems. This magneto-orbital effect is
anisotropic (i.e. depends on whether the current flow in the plane is
parallel or perpendicular to the applied parallel field direction) and
is quantitatively significant only when the parallel field is large
enough, i.e., for $l \le \langle z \rangle$ where $l=\sqrt{c\hbar/eB}$
is the magnetic length associated with the parallel field and $\langle
z \rangle$ is the approximate width of the 2D layer. In Si MOSFETs
$\langle z \rangle$ is rather small ($<50$\AA) and magneto-orbital
effects are quantitatively unimportant, making screening effect the
dominant mechanism for the parallel field dependence of the
resistivity. In both n- and p-GaAs 2D systems, both magneto-screening
and magneto-orbital effects are important, and for $B >B_s$ only the
magneto-orbital effect is operational with $\rho(B>B_s)$ increasing
monotonically due to the magneto-orbital correction. But even in GaAs
2D systems $\rho(B)$ manifests a kink at $B=B_S$ since the
spin-polarization induced screening effect saturates at that
field. The 2D parallel-field magneto-transport theory in the presence
of both magneto-spin polarization and 
magneto-orbital effects is complicated
and leads to theoretical results which are in good qualitative
agreement with experimental observations. \cite{DH7}


\section{The 2D metal-insulator transition}

We have argued that the anomalous metallic behavior of $\rho(T,n,B)$ in
2D systems arises from the qualitative variation in the effective
disorder as manifested through the screening effect. The peculiar
nature of screening in 2D (where the Thomas-Fermi screening wave
vector at $T=0$ is constant between 0 and $2k_F$, and consequently the
Kohn anomaly is a sharp cusp at $T=0$) and the attainable
low values of carrier density allow for the strong screening condition
($q_{TF}/2k_F \gg 1$) to be satisfied (for low disorder) in these
systems leading to the strong temperature and magnetic field
dependence of resistivity.

We now briefly discuss the nature of the 2D MIT itself. What is
happening at $n=n_c$? Is this a $T=0$ quantum phase transition between
a true 2D metal (not allowed within the scaling theory of
localization) and an insulator or is this a crossover (or perhaps a
classical transition) phenomenon?

It has become clear \cite{percolation,exp,lilly_p,eisenstein,yacoby}
in the last few years that the 2D MIT is a 
classical (or semiclassical) percolation transition and {\it not} a
$T=0$ quantum phase transition. The basic physical picture is simple,
and applies rather generically to semiconductor systems where the
disorder arises from a quenched {\it random distribution of Coulombic
  charged impurity centers}. At high densities the charged impurities
are effectively screened by the carrier system. As the carrier density
decreases (keeping the charged impurity density fixed, i.e. for a
given sample) the spatial 
fluctuations in the charged impurity distribution
would eventually lead to local failure (nonlinearity) in screening
leading to inhomogeneities (droplets) in the electron liquid
associated with the random impurity distribution. At low enough
carrier densities, this nonlinear screening mechanism induced
inhomogeneities would eventually lead to a percolation transition with
the effective 2D metallic phase (at $n>n_c$) above the percolation
transition point and the effective insulating phase (at $n<n_c$) 
below the percolation point. Such a percolation scenario, which should
occur quite generically in semiconductor structures where transport is
dominated by charged impurity scattering, was envisioned a long time
ago \cite{efros} for the metal-insulator transition in 3D doped
semiconductors, and has more recently been discussed for 2D systems
\cite{efros_ssc}. In fact, the percolation scenario underlying the 2D
MIT has been 
extensively discussed \cite{percolation,exp,lilly_p,eisenstein,yacoby}
in the recent literature, and there is a 
great deal of direct and indirect experimental evidence supporting the
conclusion that the 2D MIT is a percolation transition.
We mention that the same screening mechanism underlying the strong 2D
metallicity discussed in Sec. III above is responsible for the 2D MIT
phenomenon itself. At low enough carrier densities ($n\sim n_c$) the
spatial fluctuations associated with the long-range disorder potential
arising from the random charged impurity centers become too strong to
be effectively screened by the carriers, leading to screening
breakdown that provides spatial inhomogeneities (``hills'' and
``puddles'') giving rise to a percolation metal-insulator transition
in the conductivity. Thus, strong screening produces metallicity and
the nonlinear breakdown of screening produces the 2D MIT.
The crossover between the strong screening and the breakdown of
screening occurs at carrier density $n \sim n_c$, with $n_c$ being
strongly dependent on the details of the disorder in the system.

The percolation scenario also explains why weak localization effects
are difficult to observe experimentally in the 2D metallic phase. The
system is highly inhomogeneous for $n \sim n_c$, and therefore the
usual logarithmic weak localization corrections can only show up at
much higher densities where the system behaves as a disordered
homogeneous 2D electron system. Also the strong anomalous temperature
dependence arising from the screening effect makes it difficult to
directly observe the weak localization corrections to $\rho(T)$ at low
densities. The presence of strong inhomogeneities in the 2D electron
system around $n\sim n_c$ has been directly observed in scanning
chemical potential spectroscopy \cite{yacoby} of the 2D system, making
it clear 
that the 2D MIT is a screening-driven semiclassical percolation
transition rather than an interaction driven quantum phase
transition. The strong dependence of $n_c$ on disorder (e.g., on the
maximum mobility of the MOSFET) also indicates that percolation is the
underlying cause of the 2D MIT phenomenon.

It must be noted that weak localization induced negative
magnetoresistance effects are experimentally observed in the 2D
effective metallic phases even for $n \ge n_c$, i.e., just above the
2D MIT. But $\rho(T)$ itself typically saturates at low T ($\sim
30-100mK$) without manifesting the $\ln T$ rise with lowering
temperature as expected for a weakly localized 
homogeneous 2D system. This absence
of a $\ln T$ increase in $\rho(T)$ with the lowering of temperature
could be caused by a number of factors: (1) the electron heating
effect (which plays an important role below $100 mK$); (2) the strong
screening induced temperature dependence (which may mask the $\ln T$
effect); (3) inhomogeneity and droplet formation associated with
low-density nonlinear screening; (4) interaction and phase coherence
effects at low densities. It is worthwhile
to point out that even at higher carrier densities the observation of
the $\ln T$ weak localization effect in 2D semiconductor systems has
been fairly rare \cite{bishop}. More work is surely needed to better
understand 
weak localization effects in low density (and high quality) 2D systems
manifesting the 2D MIT behavior.
But the highly inhomogeneous nature of the 2D system around the
percolation transition $n\sim n_c$ suppresses the manifestation of the
weak localization behavior, which should and does show up at higher
carrier densities when the 2D system is spatially homogeneous.

We also note that the observation of a metal-insulator localization
transition goes back \cite{Ando_rmp} thirty years to the 1970s when it
was 
routinely studied in Si MOSFETs in the context of an Anderson-Mott
transition. The 2D samples used in those early studies were relatively
highly disordered samples with maximum mobilities mostly around $10^3
cm^2/Vs$ (or lower) where $n_c \approx 10^{12} cm^{-2}$ for the 2D
MIT. So the effective 2D metallic phase did not manifest any strong
metallicity since $q_{TF}/2k_F \sim 1$ and $T_F \sim 100K$. Thus, the
``transition'' to the insulating state in the pre-1994 (mostly in the
1970s) 2D MIT studies occurred without any strong metallic temperature
dependence of $\rho(T)$ at low temperatures since the effective
disorder has little temperature dependence in the $1-4K$ range for $n
\ge 10^{12}cm^{-2}$. The lack of an anomalous effective 2D metallic
phase (due to the large value of $n_c$) is what primarily
distinguishes the pre-1994 2D MIT phenomena from the post-1994 2D MIT
phenomena. 


\section{conclusion}

The low-density 2D effective metallic phase (and the eventual
lower-density 2D 
metal-insulator transition) are direct results of ohmic
low-temperature ($<1-4K$) transport in 2D semiconductor structures
being predominantly limited by long-range and singular {\it charged
  impurity} scattering (rather than by short-range white noise
disorder often used in theoretical models for the sake of
convenience). The bare disorder arising from the random charged
impurity centers is highly singular, and must be regularized for any
theoretical description of 2D transport properties. Theoretical models
that assume the quenched impurity disorder to be a short-range
white-noise potential misses out a fundamental physical 
aspect of the 2D transport properties, namely that the bare disorder
is long-range Coulomb disorder. The 2D metallicity, as
manifested in strong density, temperature, and parallel magnetic field
dependence of the resistivity (at low temperatures and carrier
densities), arises from the strong variation in the effective
(i.e. renormalized) screened disorder as a function of density,
temperature, and parallel magnetic field. The (density and temperature
dependent) screening of the charged impurity potential is {\it
  essential} in understanding the 2D transport properties -- the
situation here is fundamentally different from a short-ranged
white-noise bare disorder which does not necessitate any infrared
regularization. The dimensionless parameter $q_{TF}/2k_F$
characterizes the metallicity strength -- larger the value of
$q_{TF}/2k_F$ ($\propto n^{-1/2}$) stronger is the 2D metallic
behavior until the 2D MIT point is reached. The metallic temperature
dependence of $\rho(T,n)$ is pushed to lower absolute temperatures as
$q_{TF}/2k_F$ increases since $T_F$ (which sets the scale of
temperature variation through the dimensionless parameter $T/T_F$)
decreases with increasing $q_{TF}/2k_F$ in a given system. The 2D MIT
itself is a semiclassical percolation transition arising from the
breakdown of linear screening of the charged impurity disorder at low
enough carrier densities where the local failure of screening
associated with the randomness in the distribution of the charged
impurity centers leads to strong inhomogeneity in the electron system
(akin to electron droplet formation), eventually giving way to a
percolation transition. The critical density $n_c$ is thus a function
of the random impurity distribution, which reflects itself loosely into
a dependence of $n_c$ on the sample quality or maximum 2D mobility.

We have emphasized throughout this review that the key to
understanding the 2D MIT phenomena is that the bare disorder
here arises from random charged impurity centers, and is therefore
long-ranged and
ill-behaved. Such a singular bare Coulomb disorder {\it must be
  regularized} by screening the bare disorder. The fact that at low
carrier densities such a screened effective disorder is a strong
function of density, temperature, and parallel magnetic field is 
{\it the}
basic underlying reason for 2D metallicity and 2D MIT.

In this context, it is crucial to critique a group of recent
publications \cite{pudalov,interaction} which have confusingly attempted a
misleading and 
erroneous distinction between the screening theory 
and the so-called interaction theory \cite{zala} -- these two
theories complement each other, and are {\it not} competing
descriptions of nature. In the interaction theory \cite{zala}, one uses a
model zero-range bare impurity disorder which does not require any
regularization. The whole rationale for the screening theory does not
apply to this white-noise disorder
model, and indeed there is nothing special about the
many-body diagrams (the infinite series of ring diagrams) which define
screening when it comes to a short-range bare impurity potential. (By
contrast, the long-range Coulomb disorder, as applying to a random
distribution of charged impurity centers, must be screened, and the
infinite series of ring diagrams defining screening is a special set
of diagrams when it comes to long-range Coulombic bare impurity
disorder.) In the interaction theory, therefore, one carries out a
systematic perturbation theory in the electron-electron interaction to
obtain the {\it leading-order} temperature correction to the 2D
conductivity. Thus, the interaction theory is a formally exact
calculation of the electron-electron interaction correction to the
leading-order temperature dependence in the 2D resistivity for a model
of zero-range impurity disorder whereas the screening theory is an
approximate (the ring-diagram series) calculation of the
electron-electron correction to the {\it full} temperature dependence
in the 2D resistivity for the realistic Coulomb impurity disorder. We
emphasize that, if the dominant bare disorder in 2D semiconductor
structures were indeed some short-range quenched disorder (e.g., alloy
disorder scattering), then it would be completely meaningless to talk
about a screening theory since screening (i.e., the infinite series of
ring diagrams) has no particular significance for short range bare
disorder. On the other hand, the realistic bare disorder in 2D
semiconductor structures being the long-range Coulomb disorder,
screening is a particularly meaningful theoretical approximation for
the regularization of impurity disorder, and the assumption of a model
white-noise zero-range bare disorder makes little sense.
In this context we also mention that the Friedel oscillations
associated with impurity scattering, that are much discussed in the
interaction theory \cite{zala}, are formally equivalent to the
sharp Kohn anomaly of the screening function we described above -- the
Kohn anomaly directly leads to the Friedel oscillations.

It is important to point out that although the screening theory and
the interaction theory are complementary descriptions (and the
interaction theory does {\it not} apply to 2D semiconductor structures
since the bare disorder here is {\it not} short-range white-noise
potential), they both predict the leading-order temperature correction
to the 2D resistivity to be linear in $T/T_F$, i.e., $\rho(T) =
\rho_0[1+ f\frac{T}{T_F}]$ in both theories upto $O(T/T_F)$, i.e., for
$T/T_F \ll 1$. The parameter $f\equiv f(r_s)$ is a known function of
$r_s$ in the screening theory whereas it is an unknown function of
$r_s$ in the interaction theory where $f(r_s)$ can be expressed in
terms of various exact Fermi liquid parameters (e.g., the exact
quasiparticle effective mass and the exact quasiparticle
susceptibility) which are of course unknown functions of $r_s$ for
arbitrary $r_s$. 
The fact that both theories predict a linear
leading order temperature correction to the resistivity is highly
significant, showing the two theories to be complementary and not
competitive. The real significance of the interaction theory is
theoretical: it establishes that the metallicity (i.e. the temperature
dependence of $\rho(T)$ at low temperatures in the metallic phase)
obtained in the simple screening picture remains approximately
qualitatively valid even when higher-order interaction diagrams are
included in the theory. Another important theoretical aspect of the
interaction theory is the explicit demonstration of the connection
between the diffusive and the ballistic regimes of interaction 
temperature correction in the 2D conductivity.

It may be interesting (and certainly reasonable) to ask how
one could validate the essential nature of the screening induced
regularization of the bare Coulomb disorder in realistic 2D systems of
interest in the 2D MIT phenomena. To put the same issue in another
way: How do we know that the effective disorder is indeed the screened
Coulomb disorder in 2D semiconductor structures (as against, for
example, a zero-range white-noise disorder as assumed in the
interaction theory calculations)? This question has a surprisingly
simple experimentally verifiable answer. Writing $\rho(T,n) = \rho_0
[1 + f(r_s)\frac{T}{T_F}]$, we see that $\rho_0 \equiv \rho(T\rightarrow 0)$
is a general function of density, which is experimentally known. In
particular, at higher carrier densities ($n \gg n_c$) $\rho_0(n) \sim
n^{-1.6}$ $(n^{-1.3})$ in 2D n-GaAs (Si MOS) high mobility
semiconductor structures, which is precisely the prediction of the
screening theory for long-range bare impurity disorder
whereas the white-noise short-range disorder model
predicts $\rho_0(n) \sim n^{-1}$ independent of the 2D system. This
clearly establishes that the appropriate model for impurity disorder
is the screened Coulomb disorder, which then leads to strong
temperature and parallel magnetic field dependence of resistivity
through the strong temperature and field dependence of the screened
effective disorder for large values of $q_{TF}/2k_F$.

This dichotomy between the screening theory (which provides the full
temperature dependence of resistivity for the realistic model of
charged impurity Coulombic disorder within the simplified and
physically motivated ring diagram approximation) and the interaction
theory (which provides the formally exact leading order temperature
dependence for the unrealistic zero-range white-noise bare impurity
disorder) makes it meaningless to compare them -- it is really
comparing ``apples and oranges'' -- since the two theories serve very
different (and complementary) purposes. Where they coincide (e.g., in
the leading-order temperature correction) they provide consistent
answers. The development of a realistic interaction theory, which
starts from the realistic model of charged impurity disorder and adds
on interaction effects in a systematic manner, remains an important
open theoretical
challenge for the future. We note that this cannot be done 
within the interaction theory by
n\"{a}ively screening the bare Coulomb disorder to convert it to a
short-range effective disorder since that will be an incorrect
double-counting of the interaction effects -- the screening ring
diagrams are included in the interaction effect as the Hartree
contribution and cannot therefore be included in the regularization of
the $T=0$ bare Coulomb disorder. Finally, we note that a careful
recent experimental attempt \cite{noh} at the `verification' of the
interaction theory ran into severe consistency problems, most likely
due to the lack of information on the exact Fermi liquid parameters
necessary for applying the interaction theory to analyze experimental
data. 
We also mention that a recent theory \cite{lyo} of the thermoelectric
effect 
in 2D carrier systems has independently established the validity of
the screening theory in handling the long-range impurity disorder in
semiconductor structures.

We conclude by emphasizing that although the screening theory
``explains'' the 2D MIT phenomenon 
at a qualitative zeroth-order level, many open questions remain at
the quantitative level. In particular, the screening theory is
currently based \cite{DH1,DH2,DH3,DH4,DH5} on the ring diagram
approximation (or equivalently 
RPA, the random phase approximation) which is known to be exact only
in the high density ($r_s \rightarrow 0$) limit. Systematic inclusion
of local field corrections beyond RPA does not change the predicted 2D
metallic behavior, and in fact RPA is known to work qualitatively well
even at low carrier densities since it is a self-consistent field
theory (and {\it not} a perturbative expansion in $r_s$ which is bound
to fail at the large $r_s$ values of interest in the 2D MIT
phenomena). An important open challenge, already discussed above, is
to develop an interaction theory for the realistic long-range Coulomb
disorder in 2D systems. A deeper quantitative understanding of the
parallel magnetic field dependence of the 2D MIT behavior is also
necessary, particularly in the regime where both the spin-polarization
and the magneto-orbital effects are equally important. Much
more work is also required to better understand the precise nature of the
percolation transition itself at $n_c$. In particular, a systematic
study of the transition as a function of temperature, as has recently
been carried out \cite{lilly_p} in 2D n-GaAs system, is needed for all
the 2D 
systems in order to better understand the intriguing and interesting
2D metal-insulator transition phenomena. A better understanding of the
nature of the 2D insulating phase for $n <n_c$ is also needed.
The interplay between strong disorder and strong interaction is likely
to make the insulating phase some type of a Wigner glass or a strongly
correlated Anderson insulator, whose properties should be further
studied. We emphasize that the screening theory and the semiclassical
percolation picture are at best zeroth-order minimal descriptions for
the 2D metallicity and 2D MIT -- the important deep question about the
true nature of a 2D disordered interacting quantum ground state 
remains open.

Finally, the outstanding open theoretical question of great importance
is the nature of the true $T=0$ 2D ground state of a disordered
interacting electron system (for $n>n_c$ in particular, i.e., in the
$T\ne 0$ effective metallic phase) when both interaction and disorder
are equally significant (as they are at low carrier densities in the
presence of random charged impurity centers). 
Our current knowledge \cite{rmp} is based mostly on perturbative
techniques when either interaction or disorder (or both) are weak, and
since the system flows to the strong coupling regime when both
disorder and interaction are turned on, such perturbative theories may
not be meaningful. Direct numerical simulations \cite{kotlyar} of
rather small 2D interacting disordered systems seem to indicate that
the noninteracting scaling localization behavior holds even in the
presence of strong interaction, but thermodynamic conclusions based on
such small system simulations are open to skepticism. We emphasize in
this context that both screening theory and interaction theory, of
course, predict the $T=0$ 2D ground state to be an insulator in the
scaling localization sense, but neither theory can shed much light on
the weak localization properties in the low-density $n\ge n_c$ region
where the 2D system becomes highly inhomogeneous near the percolation
transition at $n=n_c$. While experimental work by itself is unlikely
to tell us the true nature of the $T=0$ 2D quantum ground state, we
believe that more work on carefully studying the weak localization
properties of 2D electron systems would be helpful. We note, however,
that the unmistakable evidence for the presence of weak localization
behavior in the $n>n_c$ regime is often observed experimentally
\cite{weak} although the observed localization effect seems to be
rather small, which may be due to the strong screening induced
temperature dependence and/or the inhomogeneous nature of the 2D system
around $n\ge n_c$. This issue needs to be further studied.
Limitation of space does not allow us to discuss several interesting
and important new developments in the 2D MIT physics. Recent
measurements \cite{r2} of the temperature dependence of the
weak-field Hall resistance in the 2D metallic phase show good
agreement with the screening theory \cite{r3}. Recent measurements on
the temperature dependent parallel field magnetoresistance \cite{r4}
in 2D Si/SiGe electron systems are also in good agreement with
detailed calculations based on the screening theory \cite{r5}. Because
of lack of space we are also unable to discuss several recent
theoretical developments dealing with electron-electron interaction
effects in the context of 2D metallicity and 2D MIT \cite{r6}.

\section*{ACKNOWLEDGMENTS}

This work is supported by ONR, ARO, ARDA, DARPA, LPS, NSF, and DOE.
We thank E. Abrahams, E. Demler, J. P. Eisenstein, S. M. Girvin, and
F. Stern for their comments on the manuscript.


\end{document}